# Gluonic contribution to the nucleon self-energy in an effective theory

Alka Upadhyay [1] and J.P. Singh [2]

Physics Department, Faculty of Science

M.S. University of Baroda, Vadodara-390002, India

## ABSTRACT

Gluonic contributions to the self-energy of a nucleon have been investigated in an effective theory. The couplings of the topological charge density to nucleons give rise to OZI violating η-nucleon and η'-nucleon interactions. The one-loop self-energy of a nucleon arising due to these interactions have been calculated using a heavy baryon chiral perturbation theory. The divergences have been regularized using form factors. The nontrivial structure of the QCD vacuum has also been taken into account.

[1]E-mail : alkamsub@yahoo.com

[2]E-mail : janardanmsu@yahoo.com

## 1. Introduction

The axial anomaly is known to be one of the most subtle effects of the quantum field theory. In QCD, the most important consequence of the axial anomaly is the fact that the would-be ninth Goldstone boson, the η', is massive even in the chiral limit[1]. This extra mass is induced by non-perturbative gluon dynamics[2] and the axial anomaly. The role of gluonic degrees of freedom and OZI violation in the η'- nucleon system has been investigated through, among others, the flavour –singlet Goldberger-Treiman relation[3], which, in the chiral limit, reads

$$Mg_A^{(0)} = \sqrt{\frac{3}{2}} F_0 (g_{\eta'NN} - g_{QNN}) \tag{1}$$

Here $g_A^{(0)}$ is the flavour-singlet axial-charge measured in polarised deep inelastic scattering, $g_{\eta'NN}$ is the η'-nucleon coupling constant, $g_{QNN}$ is the one-particle irreducible coupling of the topological charge density $Q = \frac{\alpha_s}{4\pi} G\tilde{G}$ to the nucleon. In Eq.(1), M is the nucleon mass and $F_0$ renormalises[4] the flavor-singlet decay constant. The coupling constant $g_{QNN}$ is, in part, related[3] to the amount of spin carried by polarized gluons in a polarized proton. The large mass of η' and the small value of $g_A^{(0)}$ (=0.2 – 0.35), extracted from deep inelastic scattering[5], point to substantial violations of the OZI rule in the flavour-singlet $J^P = 1^+$ channel[6]. A large positive $g_{QNN} \sim 2.45$ is one possible explanation of small value of $g_A^{(0)}|_{DIS}$.

It is important to look for other significant consequences which are sensitive to $g_{QNN}$. OZI violation in the η'-nucleon system is a probe of the role of the gluons in dynamical chiral symmetry breaking in low-energy QCD. It will be interesting to calculate the nucleon self-energy due to this kind of gluonic interaction. The gluonic contribution to the nucleon self-energy will be over and above the contributions associated with meson exchange models. It is known[7] that the pion self energy to the nucleon is negative, and it alone contributes (10%-20%) of the nucleon mass. Our objective in this work is to calculate self-energy due to this kind of gluonic interaction.

In the conventional chiral perturbation theory, the masses of the ground state baryon octet can be expanded in quark mass as [8] ($m_P^2 \sim m_q$):

$$M_B = M_0 + \sum_q a_q m_q + \sum_q b_q m_q^{3/2} + \sum_q c_q m_q^2 + \ldots\ldots \quad\quad\quad \ldots (2)$$

Borasoy[9] has shown that $\eta'$ can also be included in baryon chiral perturbation theory in a systematic way. In this approach, $\eta'$ is included as a dynamical field variable instead of integrating it out from the effective field theory. It has a justification in $1/N_C$ expansion where the axial anomaly is suppressed by powers of $1/N_C$ and $\eta'$ appears as a ninth Goldstone boson[2].

## 2. The low-energy effective Lagrangian

Independent of the detailed QCD dynamics, one can construct low-energy effective chiral Lagrangians which include the effect of the anomaly and axial U(1) symmetry, and use these Lagrangians to study low-energy processes involving the $\eta$ and $\eta'$ with OZI violation.

In the meson sector, the $U_A(1)$ extended low-energy effective Lagrangian can be written as[10]

$$L_{meson} = \frac{F_\pi^2}{4} Tr(\partial^\mu U \partial_\mu U^+) + \frac{F_\pi^2}{4} Tr[\chi_0(U+U^+)] + \frac{1}{2} iQ Tr[\log U - \log U^+] + \frac{3}{m_{\eta_0}^2 F_0^2} Q^2, \quad\quad \ldots(3)$$

where $U = \exp(i\frac{\phi}{F_\pi} + i\sqrt{\frac{2}{3}}\frac{\eta_0}{F_0})$ and $\phi = \sum_k \phi_k \lambda_k$ with $\phi_k$ denoting the octet of would-be Goldstone bosons($\pi, K, \eta_8$) arising out of spontaneous breaking of chiral SU(3)$_L \times$ SU(3)$_R$ symmetry. $\eta_0$ is the singlet boson and Q is the topological charge density; $\chi = diag[m_\pi^2, m_\pi^2, (2m_k^2 - m_\pi^2)]$ is the meson mass matrix, the pion decay constant $F_\pi$ =92.4 MeV and $F_0$ renormalises the flavor-singlet decay constant. The $U_A(1)$ gluonic potential involving Q is constructed to reproduce the axial anomaly in the divergence of the renormalized axial-vector current[11]:

$$\partial^\mu J_{\mu 5}^{(0)} = \sum_{k=1}^{f} 2i(m_k \overline{q_k}\gamma_5 q_k) + N_f[\frac{\alpha_s}{4\pi} G^{\mu\nu}\tilde{G}_{\mu\nu}] \quad\quad\quad \ldots(4)$$

and to generate the gluonic contribution to the η and η' masses. Here $J_{\mu 5}^{(0)} = \bar{u}\gamma_\mu \gamma_5 u + \bar{d}\gamma_\mu \gamma_5 d + \bar{s}\gamma_\mu \gamma_5 s$, $N_f$ =3, $G_{\mu\nu}$ is the gluon field strength tensor, $\tilde{G}_{\mu\nu} = \frac{1}{2}\varepsilon^{\mu\nu\alpha\beta} G_{\alpha\beta}$, and $Q(z) = \frac{\alpha_s}{4\pi} G^{\mu\nu}(z)\tilde{G}_{\mu\nu}(z)$.

The low-energy effective Lagrangian $L_{meson}$ is readily extended to include η-nucleon and η'-nucleon couplings. The chiral Lagrangian for the meson-baryon coupling upto O(p) in the meson momentum is [4]

$$L_{mb} = Tr\bar{B}(i\gamma_\mu D^\mu - M_0)B + FTr(\bar{B}\gamma_\mu \gamma_5[a^\mu, B]) + DTr(\bar{B}\gamma_\mu \gamma_5\{a^\mu, B\}) + \quad \ldots (5)$$

$$\frac{i}{3}KTr(\bar{B}\gamma_\mu \gamma_5 B)Tr(U^+\partial^\mu U) - \frac{G_{QNN}}{2M_0}\partial^\mu Q Tr(\bar{B}\gamma_\mu \gamma_5 B) + \frac{C}{F_0^4}Q^2 Tr(\bar{B}B)$$

Here B denotes the baryon octet and $M_0$ denotes the baryon mass in the chiral limit. $D_\mu = \partial_\mu - iv_\mu$ is the chiral covariant derivative, $v_\mu = -\frac{i}{2}(\xi^+\partial_\mu\xi + \xi\partial_\mu\xi^+)$ and $a_\mu = -\frac{i}{2}(\xi^+\partial_\mu\xi - \xi\partial_\mu\xi^+)$ where $\xi = U^{1/2}$. The SU(3) couplings are $F = 0.459 \pm 0.008$ and $D = 0.798 \pm 0.008$. The axial-vector current has an expansion

$$a_\mu = -\frac{1}{2F_\pi}\partial_\mu\phi - \frac{1}{2F_0}\sqrt{\frac{2}{3}}\partial_\mu\eta_0 + \ldots\ldots.$$

In continuum QCD, dynamical chiral symmetry breaking is normally studied using Dyson-Schwinger equation for quark and gluon Green's functions[12]. In low-energy effective theory given by Eqs. (3) and (5), a flavor independent self-energy of baryons will arise due to interactions of baryons with the topological charge density Q which is a flavor singlet as well as color singlet object. This gluonic term Q has no kinetic energy term, but it mixes with $\eta_0$ to generate gluonic mass term for the η'. The determination of masses of the physical η and η' mesons also requires digitalization of the ($\eta_8, \eta_0$) mass matrix. Thus, part of the η mass is also generated by the gluonic term Q[13].

The relativistic framework including baryons poses problem due to the existence of a new mass scale, namely the baryons mass in the chiral limit $M_0$; a strict chiral counting scheme, i.e., a one-to-one correspondence between the meson loops and the chiral expansion does not exist. In order to overcome this problem one integrates out the heavy degrees of freedom of the baryons, similar to a Foldy-Wouthuysen transformation, so that a chiral counting scheme emerges. Observables can then be expanded simultaneously in the Goldstone boson octet masses and the η' mass that does not vanish in the chiral limit. One obtains a one-to-one correspondence between the meson loops and the expansion in their masses and derivatives both for octet and singlet[9].

After integrating out the heavy degrees of freedom of the baryons from the effective theory[14] and assigning a four-velocity v to the baryons, the heavy baryon Lagrangian to the order we are working, reads as

$$L_{mb} = Tr(\bar{B}iv.DB) + 2FTr(\bar{B}S_\mu[a^\mu,B]) + 2DTr(\bar{B}S_\mu\{a^\mu,B\}) +$$

$$2\frac{i}{3}KTr(\bar{B}S_\mu B)Tr(U^+\partial^\mu U) - \frac{G_{QNN}}{M_0}(\partial^\mu Q)Tr(\bar{B}S_\mu B) + \frac{C}{F_0^4}Q^2 Tr(\bar{B}B) \qquad ....(6)$$

where $S_\mu = \frac{i}{2}\gamma_5\sigma_{\mu\nu}v^\nu$ is the Pauli-Lubanski spin vector.

In this work, our objective is to calculate the masses of baryon octets arising due to gluonic terms within the framework of heavy baryon chiral perturbation theory including the η'. We restrict ourselves to the one-loop diagrams of the η and η' with the vertices arising due to gluonic interactions with the baryons. For this purpose, we use the following matrix elements[15,16]:

$$\langle 0|Q|\eta\rangle = \frac{1}{\sqrt{3}}m^2_\eta(f_8\cos\theta - \sqrt{2}f_0\sin\theta)$$

$$\langle 0|Q|\eta'\rangle = \frac{1}{\sqrt{3}}m^2_{\eta'}(f_8\sin\theta + \sqrt{2}f_0\cos\theta)$$

where

$$\langle 0|J^{(8)}_{\mu5}|\eta(p)\rangle = 2if_8\cos\theta\, p_\mu, \quad \langle 0|J^{(8)}_{\mu5}|\eta'(p)\rangle = 2if_8\sin\theta\, p_\mu,$$

$$\langle 0|J^{(0)}_{\mu5}|\eta(p)\rangle = -\sqrt{6}if_0\sin\theta\, p_\mu, \quad \langle 0|J^{(0)}_{\mu5}|\eta'(p)\rangle = \sqrt{6}f_0\cos\theta\, p_\mu, \text{ and}$$

$$J^{(8)}_{\mu 5} = \frac{1}{\sqrt{3}}(\bar{u}\gamma_\mu\gamma_5 u + \bar{d}\gamma_\mu\gamma_5 d - 2\bar{s}\gamma_\mu\gamma_5 s) .$$

## 3. Regularization and the self-mass

Both the one-loop diagrams given by Figs.1(a) and 1(b) are divergent. However, we must remember that we are working in an effective field theory which uses the degrees of freedom and the interactions which are correct only at low energy. It has been shown by Donoghue et al.[8] that any incorrect loop contribution coming from short distance physics can be compensated for by a shift of the parameters of the Lagrangian. Our choice of ultraviolet regulator, which represents a separation scale of long distance physics from the short distance physics, will be dictated by phenomenological considerations. In baryon chiral perturbation theory, which deals with baryons and Goldstone bosons, the separation scale is taken as~1fm[8] corresponding to the measured size of a baryon. For our problem, we consider an average "gluonic transverse size" of nucleon[17] $<\rho^2> \approx 0.24 \text{fm}^2$ corresponding to a dipole parameterization:

$$H_g(x,t) \propto (1 - \frac{t}{m_g^2})^{-2} , \quad m_g^2 = 1.1 GeV^2, \quad x \sim 10^{-1}$$

This gives a two-gluon form factor, which we denote by u, of a nucleon [18] and can be used in the self-energy diagram. Another way to look at this problem is that the $U_A(1)$ gluonic potential involving the topological charge density leads to a contact interaction at a "short distance" (~0.2 fm) where glue is excited in the interaction region[4] of the proton-proton collision and then evolves to become an η or η' in the final state. This will lead to a sharp cutoff at an energy scale ~1GeV. In the tadpole diagram, we may use u, $u^2$, $u^{3/2}$ (geometric mean of the first two), since the phenomenology does not provide any clear-cut rule for this. Similarly, three types of form factors will be used in the tadpole diagram for exponential regularization also. In the monopole case, use of u in the tadpole diagram does not remove the divergence while $u^{3/2}$ remains analytic in a restricted region; hence we use only $u^2$. Specifically, our form factor u(k) for monopole, dipole and exponential regularization has the form:

$$u(k) = \Lambda^2/(\Lambda^2-k^2), \quad \Lambda^4/(\Lambda^2-k^2)^2, \quad \exp(k^2/\Lambda^2) .$$

Dimensional regularization scheme is not particularly suitable for effective field theories since it gets large contributions from short distance physics[8]. We have displayed our numerical results for the self-mass of the nucleon coming from both Figs.(1a,b), δm, in Table I. As discussed above, if we consider the regulator mass $\Lambda \sim$ 1GeV for the dipole and the sharp cut-off regularization schemes on phenomenological ground, we observe that δm for dipole ($u^{3/2}$-column), exponential ($u^2$-coloumn) and sharp cut-off schemes are approximately same for each mixing angle. Furthermore, δm for monopole form factor is related to that for exponential form factor (both for $u^2$-columns) by their regulator scales [7]: $\Lambda_{exp} \approx \sqrt{2}\ \Lambda_{mon}$. Hence, we take δm $\simeq$ -0.076 GeV (θ=-18.5°),

$\simeq$ -0.030 GeV (θ=-30.5°).

If we take the nontrivial structure of the QCD vacuum into account then in the last term of Eq.(6), we can make the replacement $Q^2 \to <Q^2> + Q^2$. $<Q^2>$ can be calculated using vacuum saturation hypothesis:

$<Q^2> = (-1/384) < \frac{\alpha_s}{\pi} G^2 >^2 \simeq -(1/384)(0.012)^2 GeV^8$,

where for the gluon condensate, we have used the numerical values used by ITEP group[19]. This gives a positive contribution to the nucleon mass:

δm$^{(0)}$ $\simeq$ +0.007 GeV.

Taking this into account, we get the total contribution to the nucleon mass coming from its interaction with the topological charge density δm$_{tot}$ $\simeq$ -(2.5-7.5)% of the nucleon mass. It is known that the one-loop pion contribution to the nucleon mass is δm$_{pion}$ $\simeq$ -(10-20)% of the nucleon mass[7]. Unlike δm$_{pion}$, δm$_{tot}$ is flavor independent and is same for all the members of the octet baryon family. This kind of contribution to baryon mass will not arise in models with quark-meson interaction only. It is known that the color-magnetic-field energy in the nucleon is negative[20]. We have not talked about the role of scalar and tensor gluoniums in effective field theories. In particular, scalar gluonium can give rise to Higgs-type mechanism, but this is beyond the scope of the present work.

TABLE 1

Self-energy of a nucleon, δm, arising due to its interactions with the topological charge density in monopole, dipole, exponential and sharp cut-off schemes as a function of regulator scale $\Lambda$. $\eta - \eta'$ mixing angle $\theta$ is taken as $-18.5°$. In dipole and exponential regularizations, in the tadpole diagram the form factor u (which appears at each vertex in the self-energy diagram), $u^{3/2}$ or $u^2$ has been used. Numerical values of $\Lambda$ and δm are in GeV unit.

| $\Lambda$ | Monopole | Dipole | | | Exponential | | | Sharp cut-off |
|---|---|---|---|---|---|---|---|---|
| | | u | $u^{3/2}$ | $u^2$ | u | $u^{3/2}$ | $u^2$ | |
| 0.6 | -0.033 | -0.025 | -0.015 | -0.004 | -0.050 | -0.027 | -0.017 | -0.013 |
| 0.8 | -0.078 | -0.055 | -0.036 | -0.010 | -0.112 | -0.062 | -0.040 | -0.036 |
| 1.0 | -0.148 | -0.100 | -0.071 | -0.021 | -0.206 | -0.118 | -0.078 | -0.076 |
| 1.2 | -0.248 | -0.161 | -0.121 | -0.037 | -0.334 | -0.198 | -0.135 | -0.136 |

TABLE 2

Self-energy of a nucleon, δm, as a function of regulator scale $\Lambda$ for the same form factors as in Table 1, but for $\theta = -30.5°$.

| $\Lambda$ | Monopole | Dipole | | | Exponential | | | Sharp cut-off |
|---|---|---|---|---|---|---|---|---|
| | | u | $u^{3/2}$ | $u^2$ | u | $u^{3/2}$ | $u^2$ | |
| 0.6 | -0.012 | -0.044 | -0.006 | -0.002 | -0.019 | -0.011 | -0.007 | -0.006 |
| 0.8 | -0.027 | -0.088 | -0.015 | -0.004 | -0.042 | -0.025 | -0.017 | -0.016 |
| 1.0 | -0.049 | -0.148 | -0.027 | -0.009 | -0.073 | -0.045 | -0.032 | -0.030 |
| 1.2 | -0.080 | -0.225 | -0.044 | -0.014 | -0.115 | -0.073 | -0.052 | -0.051 |

TABLE 3

Self-energy of a nucleon, δm, in dimensional regularization ($\overline{MS}$) scheme, as a function of renormalization point $\mu$.

| $\mu$ | $\theta=-18.5°$ | $\theta=-30.5°$ |
|---|---|---|
| 0.5 | -0.260 | -0.099 |
| 0.7 | -0.163 | -0.062 |
| 1.0 | -0.060 | -0.023 |

Figure Caption: Fig (1a) Self-energy diagram
                    Fig (1b) Tadpole diagram

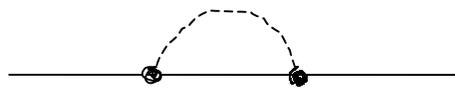

Fig (1a)

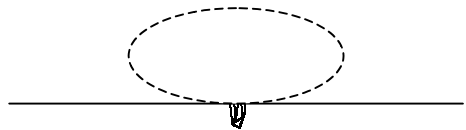

Fig (1b)